\documentclass{PoS}
\usepackage{psfig}

\PoS{PoS(LAT2005)247}

\title{Vortex-Line Percolation in the Three-Dimensional Complex Ginzburg-Landau Model\thanks{
Work supported by EU network HPRN-CT-1999-00161 EUROGRID -- ``Geometry and Disorder: from membranes
to quantum gravity'' and the Deutsche Forschungsgemeinschaft (DFG) under grant No. JA483/17-3.}}

\ShortTitle{Vortex-Line Percolation}

\author{\speaker{Elmar Bittner}, Axel Krinner and Wolfhard Janke\\
Institut f\"ur Theoretische Physik,\\
Universit\"at Leipzig,\\
Augustusplatz 10/11,\\
D-04109 Leipzig, Germany\\
        E-mail: \email{bittner@itp.uni-leipzig.de}}

\abstract{
We study the phase transition of the three-dimensional complex $|\psi|^4$ theory
by considering the geometrically defined vortex-loop network as well as the
magnetic properties of the system in the vicinity of the critical point.
Using high-precision Monte Carlo techniques we examine an alternative
formulation of the geometrical excitations in relation to the global
$O(2)$-symmetry breaking, and check if both of them exhibit the same
critical behavior leading to the same critical exponents and therefore to
a consistent description of the phase transition.
Different percolation observables are taken into account
and compared with each other. We find that different definitions of constructing
the vortex-loop network lead to different results in the thermodynamic limit,
and the percolation thresholds do not coincide with the thermodynamic
phase transition point.
}

\FullConference{XXIIIrd International Symposium on Lattice Field Theory\\
                 25-30 July 2005\\
                 Trinity College, Dublin, Ireland}

\begin{document}

Three-dimensional, globally $O(2)$ symmetric spin models and field theories
exhibit line-like topological excitations which form closed networks.
An issue of central importance is the question
whether the distribution of these vortex lines and their percolation
properties do indeed encode the critical exponents of the thermodynamically
defined phase transition. More specifically,
the question we want to address in this paper is:
Is there a similar clue in the case of vortex networks as for spin clusters,
or do they display different features?
Percolational studies of spin clusters showed that the geometric approach only works, 
if one uses a proper stochastic (Fortuin-Kasteleyn) definition of 
clusters~\cite{kasteleyn,coniglio,Fortunato,wjas}.
When connecting the vortex line elements to closed loops, which
similar to spin clusters are geometrically defined objects, and a branching point with 
$n\ge2$ junctions is encountered, a decision on how to continue has to be made. 
This step involves a certain ambiguity and gives room for a stochastic definition.
In particular we want to investigate the influence of the probability of treating such a 
branching point as a knot. 

In this paper we concentrate on the three-dimensional (3D) complex 
Ginzburg-Landau model, the field theoretical representative of the $O(2)$ 
universality class, with normalized lattice Hamiltonian 
\begin{equation}\label{h2}
H[\psi] =  \sum_{n=1}^N \Big [\frac{\sigma}{2}
(|\psi_n|^2 - 1)^2 +
\frac{1}{2}\sum_{\mu=1}^d |\psi_n-\psi_{n+\mu}|^2 \Big ]~,
\end{equation}
where $\psi_n = \psi_{n,x} + i  \psi_{n,y} = 
|\psi_n| e^{i \phi_n}$ is a complex field 
and $\sigma$ is a temperature independent parameter.
$\mu$ denotes the unit vectors along the $d$ coordinate axes, 
$N=L^d$ is the total
number of sites, and an unimportant constant term has been dropped. 
The partition function reads
\begin{equation}
Z=\int \!\! D\psi D\bar{\psi} \, e^{-H/T}~,
\label{eq:Z}
\end{equation}
where
$\int D\psi \,D\bar{\psi} \equiv \int D\,{\rm Re\/}\psi \,D\,{\rm Im\/}\psi$
stands short for integrating over all possible complex field configurations.
In the limit of a large parameter $\sigma$, it is easy to read off from 
Eq.~(\ref{h2}) that the modulus of the field is squeezed onto unity such
that  the XY model limit is approached with its well-known continuous 
phase transition in 3D at $T_c=1/\beta_c\approx 2.2$~\cite{WJ_XY}.

In order to characterize the transition we performed Monte Carlo 
simulations and measured
among other quantities the energy $\langle H\rangle$,
the specific heat $c_v=(\langle H^2\rangle- \langle H\rangle^2)/N$, 
the mean-square amplitude $\langle|\psi|^2\rangle = (1/N) \sum_{n=1}^N
\langle |\psi_n|^2\rangle$, and the magnetization $\langle M\rangle=\sum_{n=1}^N\langle\psi_n\rangle$.
The main focus of this  paper is on the properties of the geometrically 
defined vortex-loop network.
The standard procedure to calculate
the vorticity on each plaquette is by considering the quantity 
\begin{equation}
m=\frac{1}{2\pi}([\phi_1-\phi_2]_{2\pi}+[\phi_2-\phi_3]_{2\pi}+[\phi_3-\phi_4]_{2\pi}+[\phi_4-\phi_1]_{2\pi})~,
\end{equation}
where $\phi_1,\dots,\phi_4$ are the phases at the corners of a plaquette
labelled, say, according to the right-hand rule, and $[\alpha]_{2\pi}$ 
stands for $\alpha$ modulo $2\pi$: $[\alpha]_{2\pi}=\alpha+2\pi n$,
with $n$ an integer such that $\alpha+2\pi n \in (-\pi,\pi]$, hence 
$m=n_{12}+n_{23}+n_{34}+n_{41}$. If $m\neq0$, there exists a topological
charge which is assigned to the object dual to the given plaquette, i.e.,
the (oriented) line elements ${*l_\mu}$ which combine to form closed networks 
(``vortex loops''). With this definition, the vortex ``currents'' ${*l_\mu}$
can take three values: $0,\pm 1$ (the values $\pm 2$ have a 
negligible probability and higher values are impossible). The quantity
$v = \frac{1}{N}\sum_{n,\mu}|{*l}_{\mu,n}|$
serves as a measure of the vortex-line density.

In order to study percolation observables we connect the
obtained vortex line elements to closed loops.
Following a single line, there is evidently no
difficulty, but when a branching point, where $n\ge2$ junctions are encountered,
is reached, a decision on how to continue has to be made.
If we connect all in- and outgoing
line elements, knots will be formed. Another choice is to join only 
one incoming with one outgoing line element, with the outgoing
direction chosen randomly. 
We will employ two ``connectivity'' definitions here:

\begin{itemize}

\item ``Maximal'' rule: At all branching points, we 
  connect all line elements, such that the maximal loop length is
  achieved. That means each branching point is treated as a knot.

\item ``Stochastic'' rule: At a branching point where $n\ge2$
  junctions are encountered, we draw a uniformly distributed random number
  $\in (0,1]$ and if this
  number is smaller than the {\it connectivity} parameter $c$ we 
  identify this branching point as a knot of the loop,
  i.e., only with probability $0 \le c \le 1$ a branching point is treated 
  as a knot. In this way we can systematically interpolate between the
  maximal rule for $c=1$ and the case $c=0$, which corresponds 
  to the procedure most commonly followed in the literature~\cite{kajantie}.

\end{itemize}

We can thus extract from each lattice configuration a set of vortex loops,
which have been glued together by one of the connectivity definitions above. 
For each loop in the network, we measure, among others~\cite{ebakwj_PRB},
the following observables:
 
\begin{itemize}
\item ``Extent'' of a vortex loop in 1, 2, or 3 dimensions, ${\cal O}_{\rm 1D}, 
{\cal O}_{\rm 2D},$ and ${\cal O}_{\rm 3D}$:
This means simply to project the loop onto the three axes and record whether
the projection covers the whole axis, or to be more concrete, whether one
finds a vortex-line element of the loop in all planes perpendicular to the eyed
axis. 
This quantity can thus be interpreted as percolation probability~\cite{stauffer} which
is a convenient quantity for locating the percolation threshold $\beta_p$.

\item ``Susceptibilities'', $\chi_i$:
For the vortex-line density $v$ and any of the observables ${\cal O}_i$ defined 
above, one can use its variance to define the associated susceptibility,
$\chi_i=N (\langle {\cal O}_i^2 \rangle-\langle {\cal O}_i \rangle^2)$,
which is expected to signal critical fluctuations.

\end{itemize}

To update the direction of the field~\cite{hasen},
we employed the single-cluster algorithm~\cite{wolff} similar to simulations of the XY spin 
model~\cite{WJ_XY}. The modulus of $\psi$ was updated with a Metropolis algorithm. 
Here some care was necessary
to treat the measure in (\ref{eq:Z}) properly (see Ref.~\cite{ebwj_prl}).
One sweep consisted of $N$ spin flips with the Metropolis algorithm and $N_{\rm sc}$
single-cluster updates. For all simulations 
the number of cluster updates was chosen roughly proportional to the linear
lattice size, $N_{\rm sc} \simeq L$, a standard choice for 3D systems as suggested
by a simple finite-size scaling (FSS) argument.
We performed simulations for lattices with linear lattice size ranging from
$L=6$ up to $L=40$, 
subject to periodic boundary conditions.
After an initial equilibration time of $20\,000$ sweeps we took about
$100\,000$ measurements, with ten sweeps between the measurements.
All error bars are computed with the Jackknife method.

\begin{table*}[htb]
\caption{\label{expos}The critical exponents of the 3D XY model universality class
as reported in Ref.~\cite{hasen2} and the correction-to-scaling exponent $\omega$ 
of Ref.~\cite{hasen}.}
  \begin{center}
   \begin{tabular}{ccccccc}\hline\hline
    \makebox[1.6cm][c]{$\alpha$} &\makebox[1.6cm][c]{$\beta$} &\makebox[1.6cm][c]{$\gamma$} 
     &\makebox[1.6cm][c]{$\delta$}&\makebox[1.6cm][c]{$\eta$} & \makebox[1.6cm][c]{$\nu$}&\makebox[1.6cm][c]{$\omega$}\\ \hline
    $-0.0146(8)$&0.3485(2)&1.3177(5)&4.780(2)&0.0380(4)&0.67155(27)&0.79(2)\\ \hline \hline
   \end{tabular}
  \end{center}
\end{table*}
In order to be able to compare standard, thermodynamically obtained 
results (working directly with the original field variables) with the 
percolative treatment of the geometrically defined vortex-loop networks
considered here, we used the same value for the parameter $\sigma=1.5$ 
as in Ref.~\cite{ebwj_PRB} for which we determined
by means of standard FSS analyses of the magnetic susceptibility and various (logarithmic) 
derivatives of the magnetization a critical coupling of
\begin{equation}
\beta_c = 0.780\,08(4)~.
\label{eq:thermo_bc}
\end{equation}
Focussing here on the vortex loops, we performed new
simulations at this thermodynamically determined critical value, 
$\beta = 0.780\,08$, as well as additional simulations at $\beta=0.79$, $0.80$,
and $0.81$. The latter $\beta$ values were necessary because of the 
spreading of the pseudo-critical points of the vortex loop related quantities. 
As previously we recorded the time series of $H$,
$M$, $\overline{|\psi|}$, and $|\psi|^2$, as well as
the helicity modulus $\Gamma_\mu$ and the vorticity $v$. In the present 
simulations, however, we saved in addition also the field configurations in each 
measurement. This enabled us to perform the time-consuming analyses of the 
vortex-loop networks after finishing the simulations and thus to systematically 
vary the connectivity parameter $c$ of the knots.

The FSS ansatz for the pseudo-critical inverse temperatures $\beta_i(L)$, 
defined as the points where the various $\chi_i$ are maximal, is 
taken as usual as 
\begin{equation}\label{fss_beta}
\beta_i(L) = \beta_{i,c} + c_1 L^{-1/\nu}+c_2 L^{-1/\nu-\omega} + \dots ~,
\end{equation} 
where $\beta_{i,c}$ denotes the infinite-volume limit, and $\nu$ and 
$\omega$ are the correlation length and confluent correction critical 
exponents, respectively. Here we have deliberately retained the subscript $i$
on $\beta_{i,c}$.

Let us start with the susceptibility $\chi_v$ of the vortex-line density, which
plays a special role in that it is locally defined, i.e., does {\em not\/} 
require a decomposition into individual vortex loops. 
Assuming XY model values for $\nu$ and $\omega$ (cf.~Table~\ref{expos})
and fitting only the coefficients $\beta_{i,c}$ and $c_i$, we arrive at the estimate
$\beta_{v,c} = 0.779\,7(14)$
with a goodness-of-fit parameter $Q=0.20$. This value is perfectly consistent
with the previously obtained ``thermodynamic'' result (\ref{eq:thermo_bc}).
On the basis of this result one would indeed
conclude that the phase transition in the 3D complex 
Ginzburg-Landau field theory can be explained in terms of vortex-line 
{\em proliferation\/}~\cite{antunes1}.

\begin{figure*}[t]
\centerline{
\psfig{figure=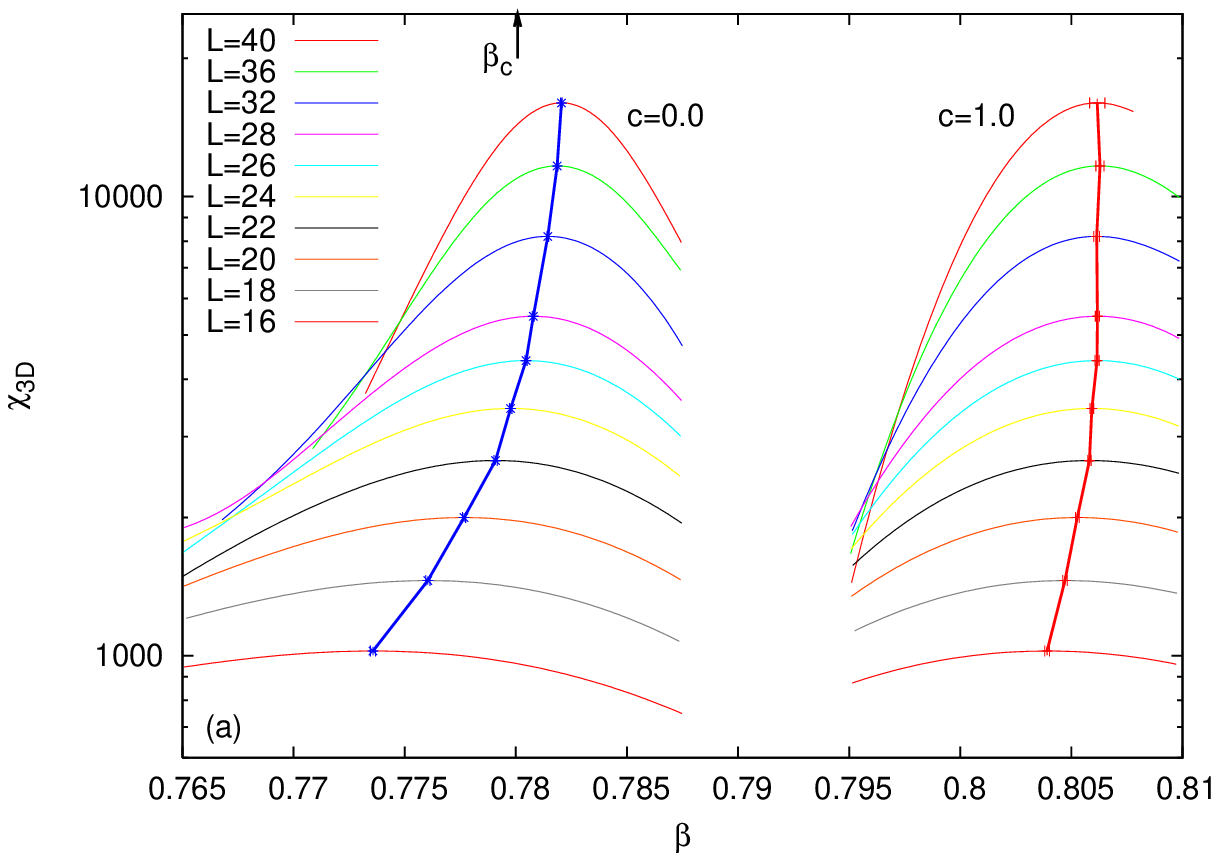,angle=0,height=5.5cm,width=7.5cm}
\psfig{figure=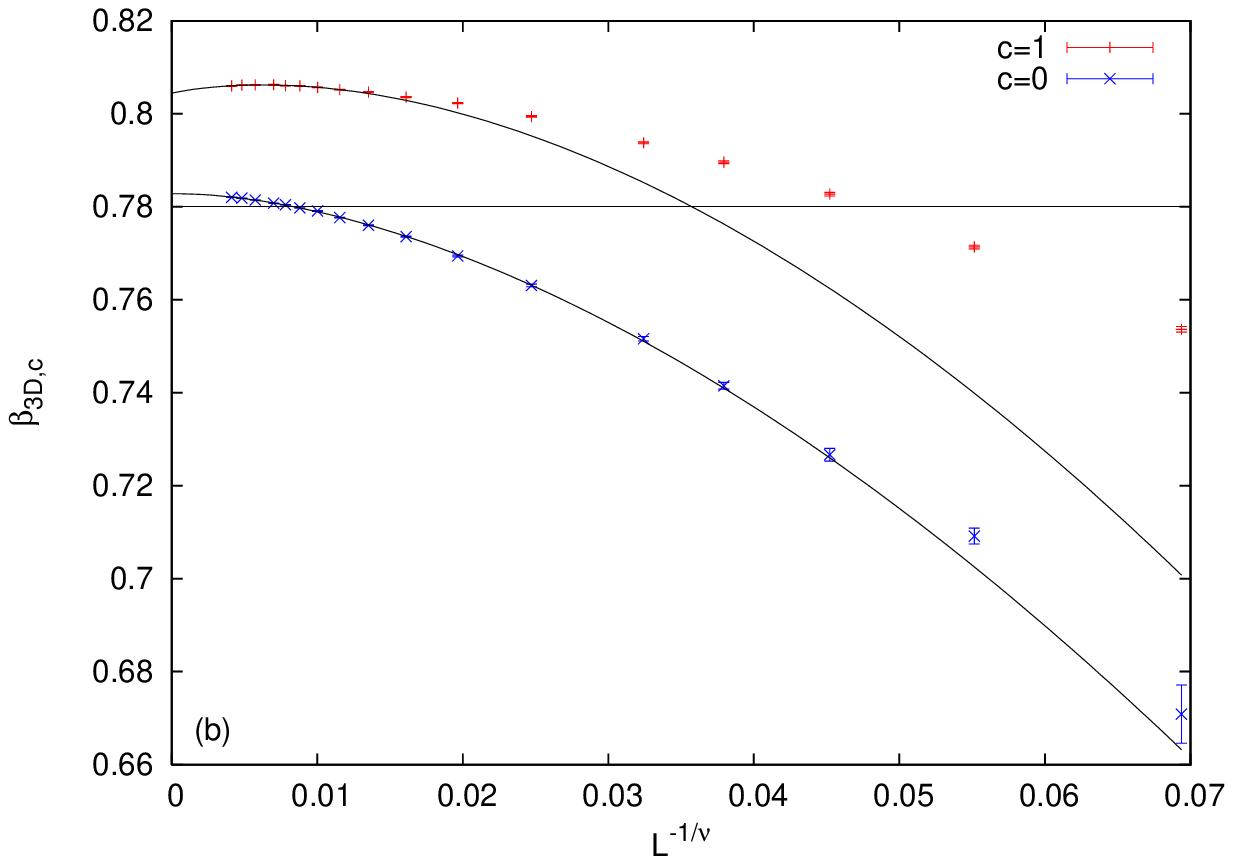,angle=0,height=5.5cm,width=7cm}
}
\caption{\label{fig:p3_1}
(a) Susceptibility $\chi_{\rm 3D}$ of ${\cal O}_{\rm 3D}$ 
as a function of $\beta$ for the ``stochastic'' ($c=0$) and the ``maximal'' ($c=1$) rule. 
(b) Location of the percolation thresholds determined from the maximum of $\chi_{\rm 3D}$ as a 
function of $L^{-1/\nu}$. The lines indicate fits using the FSS ansatz (\protect\ref{fss_beta})
with $\nu$ and $\omega$ fixed according 
to Table~\protect\ref{expos}. The horizontal dashed line shows the thermodynamically
determined critical coupling $\beta_c = 0.780\,08(4)$.}
\end{figure*}

To develop a purely geometric picture of the mechanism governing this transition,
however, one should be more ambitious and also consider the various quantities
$\mathcal{O}_i$ introduced above that focus on the {\em percolative\/} 
properties of the vortex-loop network.
As an example for the various susceptibilities considered, we show in 
Fig.~\ref{fig:p3_1}(a) the susceptibility $\chi_{\rm 3D}$ of ${\cal O}_{\rm 3D}$
for $c=0$ and $c=1$.
The scaling behavior of the
maxima locations $\beta_{\rm 3D}(L)$ of the susceptibility $\chi_{\rm 3D}$ of ${\cal O}_{\rm 3D}$
for $c=0$ and $c=1$ is depicted in Fig.~\ref{fig:p3_1}(b), where the lines indicate fits according to 
Eq.~(\ref{fss_beta}) with exponents fixed again according to Table~\ref{expos}.
We obtain $\beta_{{\rm 3D},c}=0.7824(1)$ with $\chi^2/{\rm dof}=1.14$ 
$(Q=0.32, L\ge 8)$ for $c=0$ and $\beta_{{\rm 3D},c}=0.8042(4)$ with
$\chi^2/{\rm dof}=0.75$ $(Q=0.58, L\ge 20)$ for 
$c=1$. While for the ``stochastic'' rule with $c=0$ the infinite-volume limit 
of  $\beta_{\rm 3D}(L)$ is at least close to $\beta_c$, it is clearly
significantly larger than $\beta_c$ for the fully knotted vortex networks with 
$c=1$.

By repeating the fits for all vortex-network observables and the parameter
$c$ between 0 and 1 in steps of 0.1, 
we observed that the location of the infinite-volume limit $\beta_{i,c}$ does depend on
the connectivity parameter $c$ used in constructing the vortex loops in 
a statistically significant way. 
With decreasing $c$, the infinite-volume 
extrapolations come closer towards the thermodynamical critical value 
(\ref{eq:thermo_bc}), but even for $c=0$ they clearly do not coincide.

\begin{figure*}[t]
\centerline{\psfig{figure=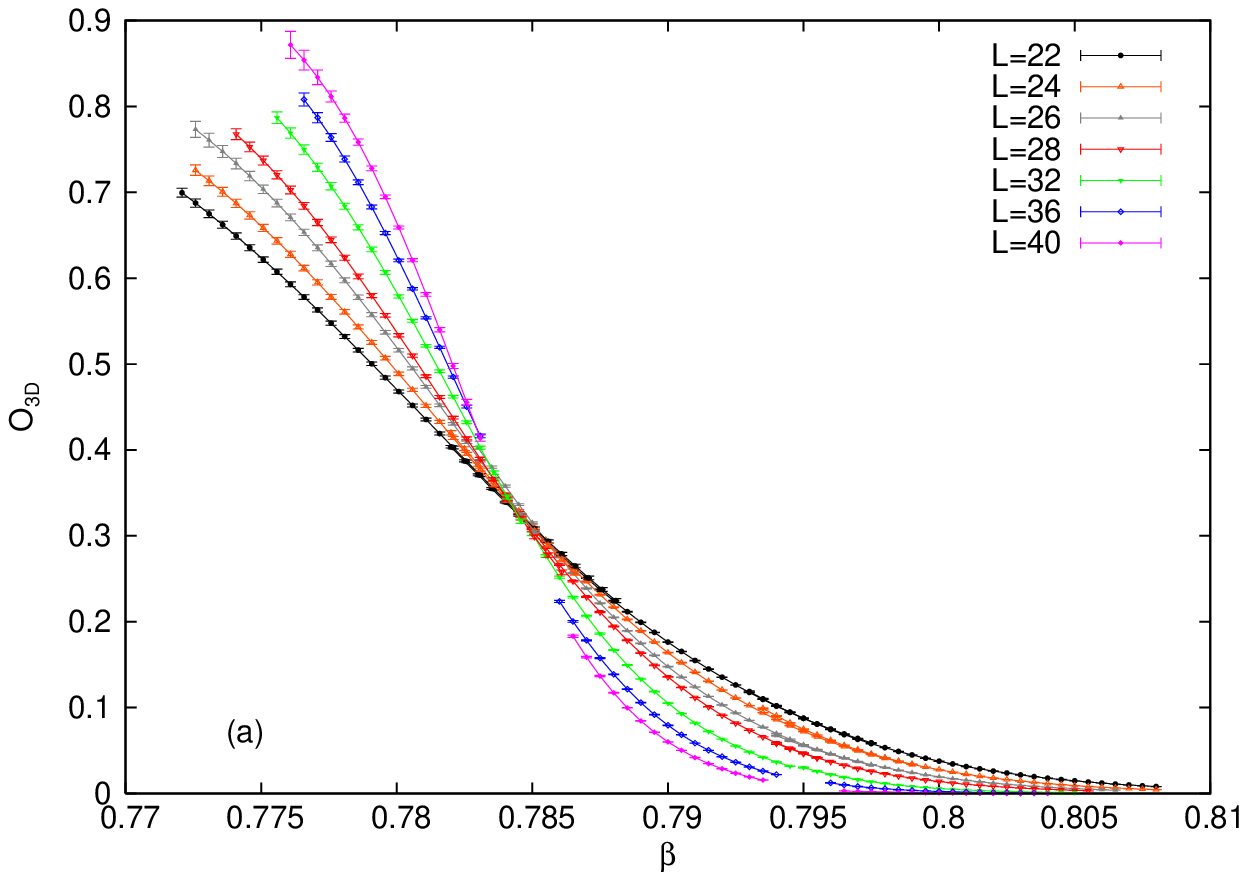,height=5.5cm,width=7cm}
\psfig{figure=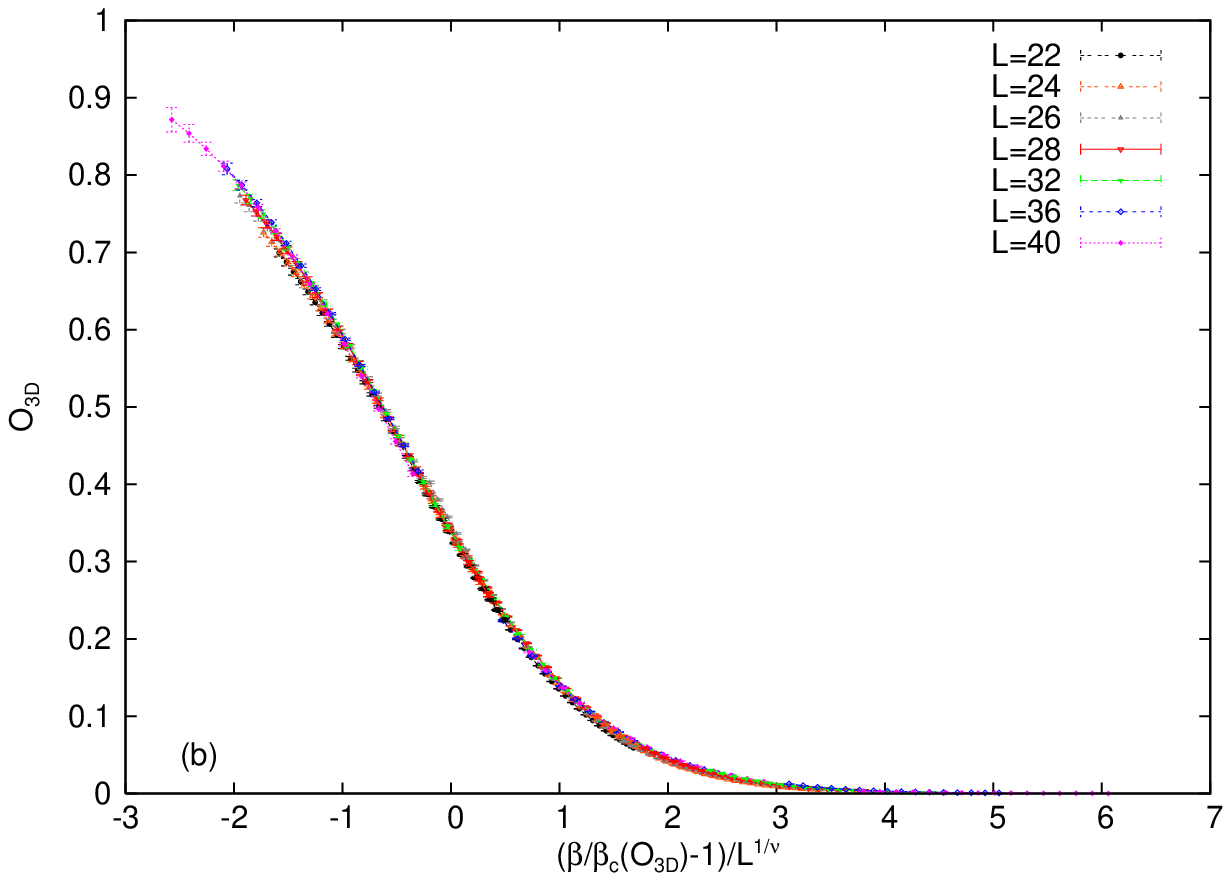,height=5.5cm,width=7cm}}  
\caption{ \label{res_O3}
(a) ${\cal O}_{\rm 3D}$ as function of $\beta$ for $c=0$. 
(b) Rescaled data with $\nu$ fixed at the 3D XY model value 
(cf.~Table~\protect\ref{expos}) and choosing $\beta_c({\cal O}_{\rm 3D})=0.7842$ from
the location of the crossing point in (a) for the best data collapse.
}
\end{figure*}

We nevertheless performed tests whether at least 
for $c=0$ the critical behavior of the vortex-loop network may consistently
be described by the 3D XY model universality class. 
As an example for a quantity that is {\em a priori\/}
expected to behave as a percolation probability we picked the
quantity $\mathcal{O}_{\rm 3D}$.
As is demonstrated in Fig.~\ref{res_O3}(a)
for the case $c=0$, by plotting the
raw data of $\mathcal{O}_{\rm 3D}$ as a function of $\beta$ for the various
lattice sizes, one obtains a clear crossing point so that the interpretation
of $\mathcal{O}_{\rm 3D}$ as percolation probability is 
nicely confirmed. To test the scaling behavior we rescaled
the raw data in the FSS master plot shown in Fig.~\ref{res_O3}(b), where the 
critical exponent $\nu$ has the XY model value given in Table~\ref{expos} and
$\beta_c (\mathcal{O}_{\rm 3D})=0.7842$ was independently determined by 
optimizing the data collapse, i.e., virtually this is the location of the
crossing point in Fig.~\ref{res_O3}(a). The collapse turns out to be quite
sharp. 
For $c>0$ we found also a sharp data collapse, but for
a monotonically increasing exponent $\nu$, which is for large $c$ values
compatible with the critical exponent $\nu=0.8765(16)$ of 3D
percolation~\cite{ballesteros}.
One should keep in mind, however, that neither 
$\beta_{{\rm 3D},c}$ as extrapolated from the susceptibility
peaks nor the estimate obtained from the crossing point in Fig.~\ref{res_O3}(a) is
compatible with $\beta_c$.

To summarize, we have found for the 3D complex
Ginzburg-Landau field theory that the geometrically defined percolation 
transition of the vortex-loop network is close to the thermodynamic phase 
transition, but does not quite coincide with it for any observable we have considered.
Our results for the connectivity parameter $c \in [0,1]$ extend the claim
of Ref.~\cite{kajantie} for the 3D XY spin model that 
neither the ``maximal'' ($c=1$) nor the
``stochastic'' rule ($c=0$) used for constructing macroscopic vortex
loops does reflect the properties of the true phase transition in a strict 
sense.
A more detailed presentation of these and additionally results for several other
observables is given in Ref.~\cite{ebakwj_PRB}.

\begin{thebibliography}{99}
\bibitem{kasteleyn}
P.W. Kasteleyn and C.M. Fortuin, \emph{J. Phys. Soc. of Japan} {\bf 26} (Suppl.) (1969) 11;
C.M. Fortuin and P. W. Kasteleyn, \emph{Physica} {\bf 57} (1972) 536;
C.M. Fortuin, \emph{Physica} {\bf 58} (1972) 393; {\em ibid.\/} {\bf 59} (1972) 545.
\bibitem{coniglio} A. Coniglio and W. Klein, \emph{J. Phys.} {\bf A13} (1980) 2775.
\bibitem{Fortunato} S.~Fortunato, \emph{J. Phys.} {\bf A36} (2003) 4269, [\href{http://xxx.lanl.gov/abs/hep-lat/0207021}{\tt hep-lat/0207021}].
\bibitem{wjas} W. Janke and A.M.J. Schakel, \emph{Nucl. Phys.} {\bf B700} (2004) 385, [\href{http://xxx.lanl.gov/abs/cond-mat/0311624}{\tt cond-mat/0311624}] .
\bibitem{WJ_XY} W. Janke, \emph{Phys. Lett.} {\bf A148} (1990) 306.
\bibitem{kajantie} K. Kajantie, M. Laine, T. Neuhaus, A. Rajantie, 
and K. Rummukainen, \emph{Phys. Lett.} {\bf B482} (2000) 114, [\href{http://xxx.lanl.gov/abs/hep-lat/0003020}{\tt hep-lat/0003020}].
\bibitem{ebakwj_PRB} E. Bittner, A. Krinner, and W. Janke, \emph{Phys. Rev.} {\bf B72} (2005) 094511. 
\bibitem{stauffer} D. Stauffer and A. Aharony, {\em  Introduction to Percolation 
Theory\/}, 2nd ed., Taylor and Francis, London, 1994.
\bibitem{hasen} M. Hasenbusch and T. T\"or\"ok, \emph{J. Phys.} {\bf A32} (1999) 6361, [\href{http://xxx.lanl.gov/abs/cond-mat/9904408}{\tt cond-mat/9904408}].
\bibitem{wolff} U. Wolff, \emph{Phys. Rev. Lett.} {\bf 62} (1989) 361;
\emph{Nucl. Phys.} {\bf B322} (1989) 759.
\bibitem{ebwj_prl} E. Bittner and W. Janke, \emph{Phys. Rev. Lett.} {\bf 89} (2002) 130201.
\bibitem{ebwj_PRB} E. Bittner and W. Janke, \emph{Phys. Rev.} {\bf B71} (2005) 024512, [\href{http://xxx.lanl.gov/abs/cond-mat/0501468}{\tt cond-mat/0501468}].
\bibitem{hasen2} M. Campostrini, M. Hasenbusch, A. Pelissetto, P. Rossi,
and E. Vicari, \emph{Phys. Rev.} {\bf B63} (2001) 214503, [\href{http://xxx.lanl.gov/abs/cond-mat/0010360}{\tt cond-mat/0010360}].
\bibitem{antunes1} N.D. Antunes, L.M.A. Bettencourt, and M. Hindmarsh,
\emph{Phys. Rev. Lett.} {\bf 80} (1998) 908, [\href{http://xxx.lanl.gov/abs/hep-ph/9708215}{\tt hep-ph/9708215}];
N.D. Antunes and L.M.A. Bettencourt, \emph {Phys. Rev. Lett.} {\bf 81} (1998) 3083, [\href{http://xxx.lanl.gov/abs/hep-ph/9807248}{\tt hep-ph/9807248}].
\bibitem{ballesteros}
H.G. Ballesteros, L.A. Fernandez, V. Martin-Mayor, A. Munoz-Sudupe, G. Parisi, and J. J. Ruiz-Lorenzo,
\emph{J. Phys.} {\bf A32} (1999) 1, [\href{http://xxx.lanl.gov/abs/cond-mat/9805125}{\tt cond-mat/9805125}].
\end{thebibliography}
\end{document}